\documentclass{aa}

\usepackage{natbib}
\bibpunct{(}{)}{;}{a}{}{,}
\usepackage{color}
\definecolor{red}{cmyk}{0,1,1,1}

\usepackage{graphicx}
\usepackage{txfonts}
\usepackage{hyperref}
\begin{document}

        \title{Biosignatures of the  Earth I. Airborne spectropolarimetric detection of photosynthetic life}

        \author{C.H. Lucas Patty
                \inst{1}
                \and
                Jonas G. K\"uhn\inst{2,3}
                \and
                Petar H. Lambrev\inst{4}
                \and
                Stefano Spadaccia\inst{1}
                \and            
                H. Jens Hoeijmakers\inst{5}
                \and
                Christoph Keller\inst{6}
                \and
                Willeke Mulder\inst{6}
                \and
                Vidhya Pallichadath\inst{7}
                \and
                Olivier Poch\inst{8}
                \and
                Frans Snik\inst{6}
                \and
                Daphne M. Stam\inst{7}
                \and
                Antoine Pommerol\inst{1}
                \and
                Brice O. Demory\inst{2}
                }

        \institute{Physikalisches Institut, Universit\"at Bern, CH-3012 Bern, Switzerland\\
                \email{lucas.patty@space.unibe.ch, chlucaspatty@gmail.com}
                \and
                Center for Space and Habitability, Universit\"at Bern,  CH-3012 Bern, Switzerland
                \and
                D\'epartement d'Astronomie, Universit\'e de Gen\`eve, CH-1290 Versoix, Switzerland
                \and
                Institute of Plant Biology, Biological Research Centre Szeged, Temesvári krt. 62, 6726 Szeged, Hungary 
                \and
                Lund Observatory, Lund University, P.O. Box 43, SE-22100 Lund, Sweden
                \and
                Leiden Observatory, Leiden University, P.O. Box 9513, 2300 RA Leiden, The Netherlands
                \and
                Faculty of Aerospace Engineering, Delft University of Technology, Kluyverweg 1, 2629 HS Delft, The Netherlands
                \and
                Universit\'e Grenoble Alpes, CNRS, IPAG, 38000 Grenoble, France
        }
        
        
        \abstract
        {Homochirality is a generic and unique property of life on Earth and is considered a universal and agnostic biosignature. Homochirality induces fractional circular polarization in the incident light that it reflects. Because this circularly polarized light can be sensed remotely, it can be one of the most compelling candidate biosignatures in life detection missions. While there are also other sources of circular polarization, these result in spectrally flat signals with lower magnitude. Additionally, circular polarization can be
a valuable tool in Earth remote sensing because\ the circular polarization signal directly relates to vegetation physiology.}
        {While high-quality circular polarization measurements can be obtained in the laboratory and under semi-static conditions in the field, there has been a significant gap to more realistic remote sensing conditions.}
        {In this study, we present sensitive circular spectropolarimetric measurements of various landscape elements taken from a fast-moving helicopter.}
        {We demonstrate that during flight, within mere seconds of measurements, we can differentiate (S/N>5) between grass fields, forests, and abiotic urban areas. Importantly, we show that with only nonzero circular polarization as a discriminant, photosynthetic organisms can even be measured in lakes.}
        {Circular spectropolarimetry can be a powerful technique to detect life beyond Earth, and we emphasize the potential of utilizing circular spectropolarimetry as a remote sensing tool to characterize and monitor in detail the vegetation physiology and terrain features of Earth itself.}
        

        \maketitle

\section{Introduction}

Terrestrial biochemistry is based on chiral molecules which, in their most simple form, occur in a left-handed (L-) and right-handed (D-) version and are non-superimposable mirror images of each other. Unlike abiotic chemistry, where these molecules occur in roughly equal concentrations of both versions, living nature almost  exclusively \citep{Grishin2020} utilizes these molecules in only one of the mirror image configurations. This is called homochirality. 

Homochirality also manifests itself within biological macromolecules and biomolecular architectures. The $\alpha$-helix, which is the most prevalent secondary structure of proteins, is almost exclusively right-hand-coiled \citep{Novotny2005}. The biochemical homochirality is essential for processes ranging from enzymatic functioning to self-replication. The latter also implies the prerequisite of homochirality for life as we know it \citep{Popa2004, Bonner1995, Jafarpour2015}. Homochirality is likely a universal feature of life \citep{Wald1957,MacDermott2012,Patty2018a,Sparks2009}; and owing to its high exclusivity, homochirality may serve as an agnostic trait of life itself.

The molecular dissymmetry of chiral molecules and architectures endows them with a specific response to electromagnetic radiation \citep{Fasman2013,Patty2018a}. This response is most evident upon interaction with polarized light, but also modifies the polarization properties of incident unpolarized light. Circular dichroism is the differential absorbance of left- and right-handed circularly polarized incident light, which only yields a net signal when the molecules are homochiral. Circular polarizance is related to circular dichroism; circular polarizance is the fractionally induced circular polarization resulting from the interaction with unpolarized incident light, such as is essentially emitted from the Sun \citep{Kemp1987}. It has been demonstrated that circular polarizance carries the same information as circular dichroism \citep{Patty2017,Patty2018b}, but unlike the latter, cicular polarizance can be sensed remotely (i.e., without controlling the source) \citep{Pospergelis1969, Wolstencroft2004,Sparks2009}.

Remote sensing of the Earth's surface with spectropolarimetry could offer information not accessible with the reflectance alone. The circular polarization response of vegetation, albeit small and typically less than 1\% around the chlorophyll absorbtion band, is an extremely sensitive indicator of the membrane macro-organization of the photosynthetic apparatus and its functioning \citep{Garab2009,Garab1996,Lambrev2019}. The polarization spectra could, in principle, be used to detect various physiological stress responses and could provide sensitive monitoring of vegetation physiology \citep{Lambrev2019}. It has for instance been demonstrated that circular polarization can be more indicative of stress factors such as drought than the scalar reflectance \citep{Patty2017}. Other phototrophic organisms can also yield clear circular polarization signals and some of these can produce relatively large signals: up to 2\% for certain multicellular algae \citep{Patty2018c}. Additionally, nonzero polarization signals can be observed for various (an)oxygenic phototrophic microorganisms and microbial mats \citep{Sparks2009a,Sparks2009,Sparks2021}. Polarization remote sensing could prove to be a very powerful complementary tool in assessing, for example  the effects of climate change, desertification, deforestation, and the monitoring of vegetation physiological state in vulnerable regions.

As a result of its dependence on homochirality and the three-dimensional molecular asymmetry of life, the circular polarization induced by life is a serious contender to be a generic agnostic (universal) surface biosignature. While most research efforts on remotely detectable biosignatures have focused on the spectroscopic characterization of planetary atmospheres, detection of gases related to the possible presence of a planetary biosphere is not free of false-positive scenarios \citep{Meadows2018a, Schwieterman2016, Schwieterman2018}. Surface biosignatures, such as circular polarization, are a direct observable characteristic of life and as such have a potential to reduce the model dependence \citep{Schwieterman2018}. While abiotic materials can also create circular polarization, such as through multiple scattering by clouds or aerosol particles, these interactions result in a weaker and spectrally flat circular polarization signal \citep{Rossi2018}. {\color{red}Additionally, incident unpolarized light cannot become circularly polarized by anisotropic Rayleigh scattering alone, but it can become circularly polarized if scattered at least once by a Rayleigh scatterer and consequently by a large aerosol or particle \citep{Rossi2018}. Furthermore, as the direction of circular polarization is directly determined by the chirality of the molecular assembly, the signal is not averaged out in a planetary disk average. Averaging out does occur for the circular polarization signal resulting from scattering by clouds or aerosol particles \citep{Rossi2018}.}

Additionally, circular spectropolarimetric measurements on various minerals, artificial grass fields, and on the surface of Mars, and the absence of nonzero signals measured in these studies attest to the general lack of false positive scenarios and abiotic mimicry \citep{Sparks2009,Sparks2009a, Pospergelis1969, Sparks2012, Patty2019}. Other suggested remotely detectable (surface) biosignatures, such as the vegetation red edge or other pigment signatures \citep{Seager2005,Kiang2007, Schwieterman2015}, have a higher risk of possible false positives by mineral reflectance.

Research on circularly polarized biosignatures or possible remote sensing applications have been developing quickly in recent years. Previous works have demonstrated the potential of high-sensitive circular polarization measurements obtained in the laboratory \citep{Patty2017, Patty2018c, Patty2018b, Sparks2009,Sparks2009a,Sparks2021,Martin2010,Martin2016,Martin2020} and in the field \citep{Patty2019}. All of these measurements, however, were taken under stable conditions in the laboratory or are measurements of semi-static (allowing integration over longer durations) scenes in the field.

In the present study we show, for the first time, sensitive circular spectropolarimetry from a fast-moving aerial platform using the spectropolarimeter FlyPol, based on the TreePol design \citep{Patty2017,Patty2018a,Patty2019}. We report the observation of circularly polarized signals for various landscape features, such as forests and grass fields, and we demonstrate that these signals can be readily distinguished from abiotic scenes such as roads and urban area merely using nonzero circular polarization spectral features as a discriminator. 

\section{Materials and methods}

Circular spectropolarimetric measurements were carried out using FlyPol. FlyPol is an adaptation of the TreePol instrument (see \citep{Patty2017,Patty2018a,Patty2019} for more information), which was specifically designed to measure the fractionally induced circular polarization ($V/I$) as a function of wavelength (400–900 nm) at high sensitivity ($<10^{-4}$) and accuracy ($<10^{-3}$). For FlyPol we upgraded the spectrographs (Avantes, The Netherlands) and amended the design with active temperature control of the optics (Thorlabs, USA) and electronics, which greatly improved the temporal stability of the instrument. An additional feature is the possibility of mechanically actuating optical elements, such as inserting and removing optical elements, which will facilitate the sequential acquisition of full-Stokes measurements in the future. The angular field of view of FlyPol is approximately  1.2$^\circ$. The instrument was oriented using a calibrated, parallax-free telescope pointer mounted on top of the instrument casing (53x33x13cm).

All the measurements were taken from an Enstrom 280C helicopter and were carried out in the districts of Val-de-Travers and Le Locle of the canton of Neuch\^atel, Switzerland. These districts lie in the Swiss Jura, which is part of the Alpine foreland and is characterized by a hilly terrain and a series of sub-parallel valleys and ridges. Measurements were taken with an angle to the vertical of approximately 45$^\circ$, at a ground speed of 20-39 kts (10-20 m/s) and up to a altitude of 5500 ft AMSL (1.7 km). A GPS receiver was used to track the flight trajectories. {\color{red} All measurements were performed under clear sky conditions on September 17, 2019 at 1600-1620 hr. During the flight, the solar zenith and azimuth angle were approximately $57^\circ$ and $233^\circ$, respectively. The measurements, however, were averaged over different phase angles, which could not be recorded accurately.} Instrumental calibration was performed on a spectralon target on the ground prior and post flight. The reflectance data displayed are qualitative and merely shown as support for the polarization spectra. The integration time for all measurements was 12 ms to prevent detector saturation. For several scenes, the average photon count (<10000 counts) was too low and these measurements were discarded.

Laboratory measurements of vegetation (\textit{Graminae sp., Acer platanoides, Platanus sp.}) and concrete were taken in reflection using an integrating sphere. For the polarimetric spectra of vegetation, three different leaves or grass patches per species were measured. The measurements on concrete were taken from three different regions on the same slab. The results were averaged over a 30 s measurement per sample. 
\begin{figure}[!ht]
        \centering 
        \includegraphics[width=0.49\textwidth]{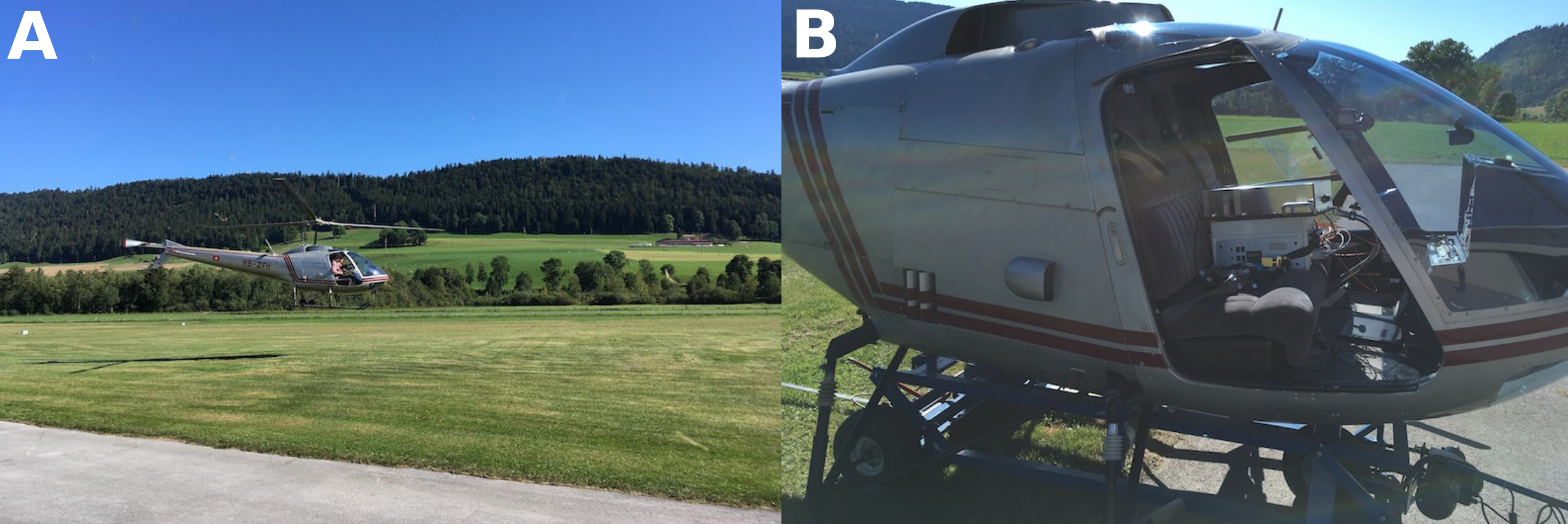}
        \caption{ Enstrom 280C taking off at A\'erodrome de M\^otiers (\textbf{A}) and a closeup of the cockpit with FlyPol inside (\textbf{B}).}
        \label{fig:Helipic}
\end{figure}

\section{Results}
\subsection{Landscape element differentiation}
\begin{figure*}[!htbp]
        \centering 
        \includegraphics[width=0.98\textwidth]{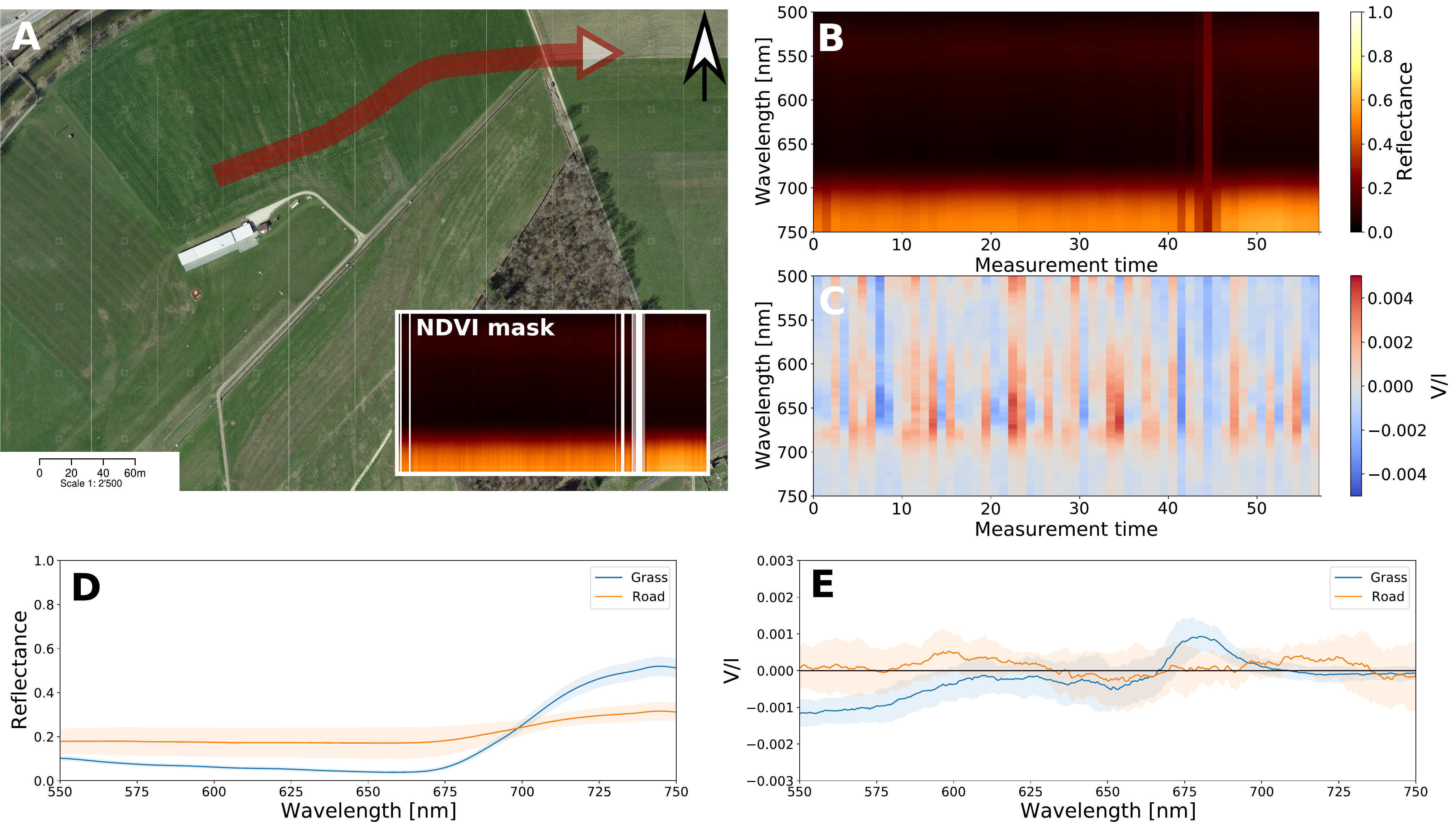}
        \caption{\textbf{A}: One of the measurement trajectories (in red) during a flight near A\'erodrome de M\^otiers, M\^otiers, flown in E direction. The inset indicates the vegetation classification using NDVI where the white is classified as non-grass (same color bar and scale as \textbf{B}). Satelite image by Swiss Federal Authorities, CNES; Spot Image, Swisstopo, NPOC. \textbf{B}: Qualitative reflectance data for the trajectory per wavelength in time [s]. \textbf{C}: V/I for the trajectory per wavelength in time [s]. \textbf{D}: Time-averaged reflectance of the grass and road, respectively. \textbf{E}: Time-averaged V/I of the grass and road, respectively. Shaded areas denote the standard deviation between measurements.}
        \label{fig:Sat}
\end{figure*}

In order to differentiate between the different landforms and landscape elements we used the differences in reflectance as measured by FlyPol and we calculated the normalized difference vegetation index (NDVI) \citep{RouseJr1973} as follows:

\begin{equation}
NDVI = \frac{I_{FR}-I_{R}}{I_{FR}+I_{R}}
,\end{equation}

where $I_{R}$ is the mean of the reflectance between 650 nm - 680 nm and $I_{FR}$ is the mean between 750 nm - 780 nm. A  comparison with the flight trajectories showed that using the NDVI was accurate in differentiating between the elements per scene. 

In Figure \ref{fig:Sat} we show the results of a trajectory recorded near A\'erodrome de M\^otiers. The approximate trajectory is shown in Figure \ref{fig:Sat} \textbf{A}, where the inset shows the mask based on the NDVI data used to filter out the two intersecting roads. Figure \ref{fig:Sat}  \textbf{B} and \textbf{C} show the reflectance and circular polarization, respectively, binned to one second measuring steps. Interestingly, while the major part of the trajectory in Figure \ref{fig:Sat} (\textbf{B}) exhibits a red edge, a circular polarization signal is not continuously detected (\textbf{C}). However, by averaging over the whole scene, a clear signal can be obtained. The results averaged over the whole scene per different landscape element, grass, and roads are shown in Figure \ref{fig:Sat} \textbf{D} and \textbf{E} for the reflectance and $V/I$. 

\begin{figure}[!htb]
        \centering 
        \includegraphics[width=0.49\textwidth]{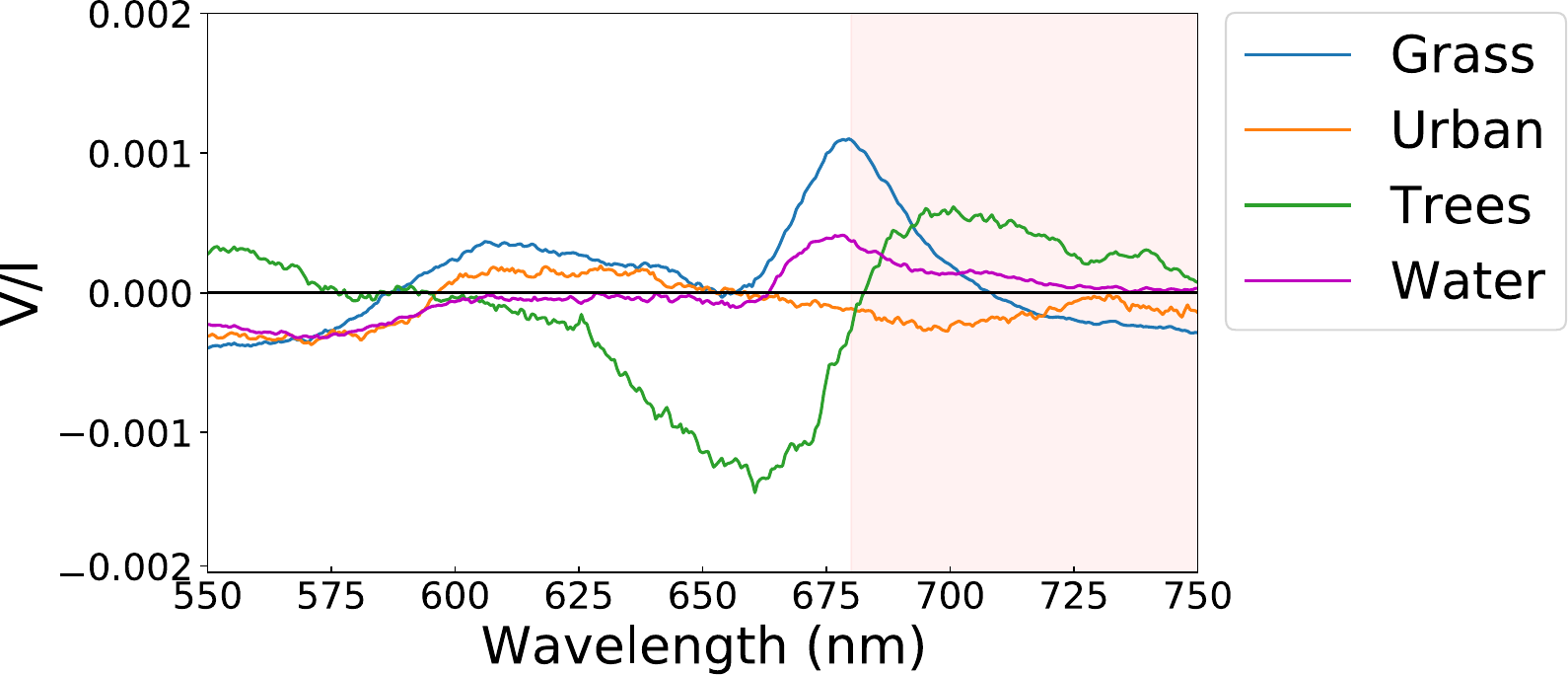}
        \caption{Circular polarization spectra for various landscape features: grass, urban area, trees, and water containing algae. The red shaded area represents the  red edge.}
        \label{fig:class}
\end{figure}

Using the NDVI, we were able to separate four different landscape elements: grass, trees, urban, and water. The resulting average circular polarization spectra for these categories are shown in Figure \ref{fig:class}. It is clearly visible that the average over the urban area has a very low signal, where $V/I_{max}=2.6\cdot10^{-4}$ at $\sim$700 nm. For water containing algae and grass, which both show a positive band with a maximum at $\sim$680 nm, the magnitudes are $V/I_{max}=4.1\cdot10^{-4}$ and $V/I_{max}=1.1\cdot10^{-3}$, respectively. The average taken over the trees shows a positive maximum of $V/I_{max}=6.1\cdot10^{-4}$ at $\sim$ 700 nm and a negative band with a value of $V/I_{min}=-1.4\cdot10^{-3}$ at $\sim$660 nm.

\subsection{Lacustrine aerial spectropolarimetry}
\begin{figure*}[!htb]
        \centering 
        \includegraphics[width=1\textwidth]{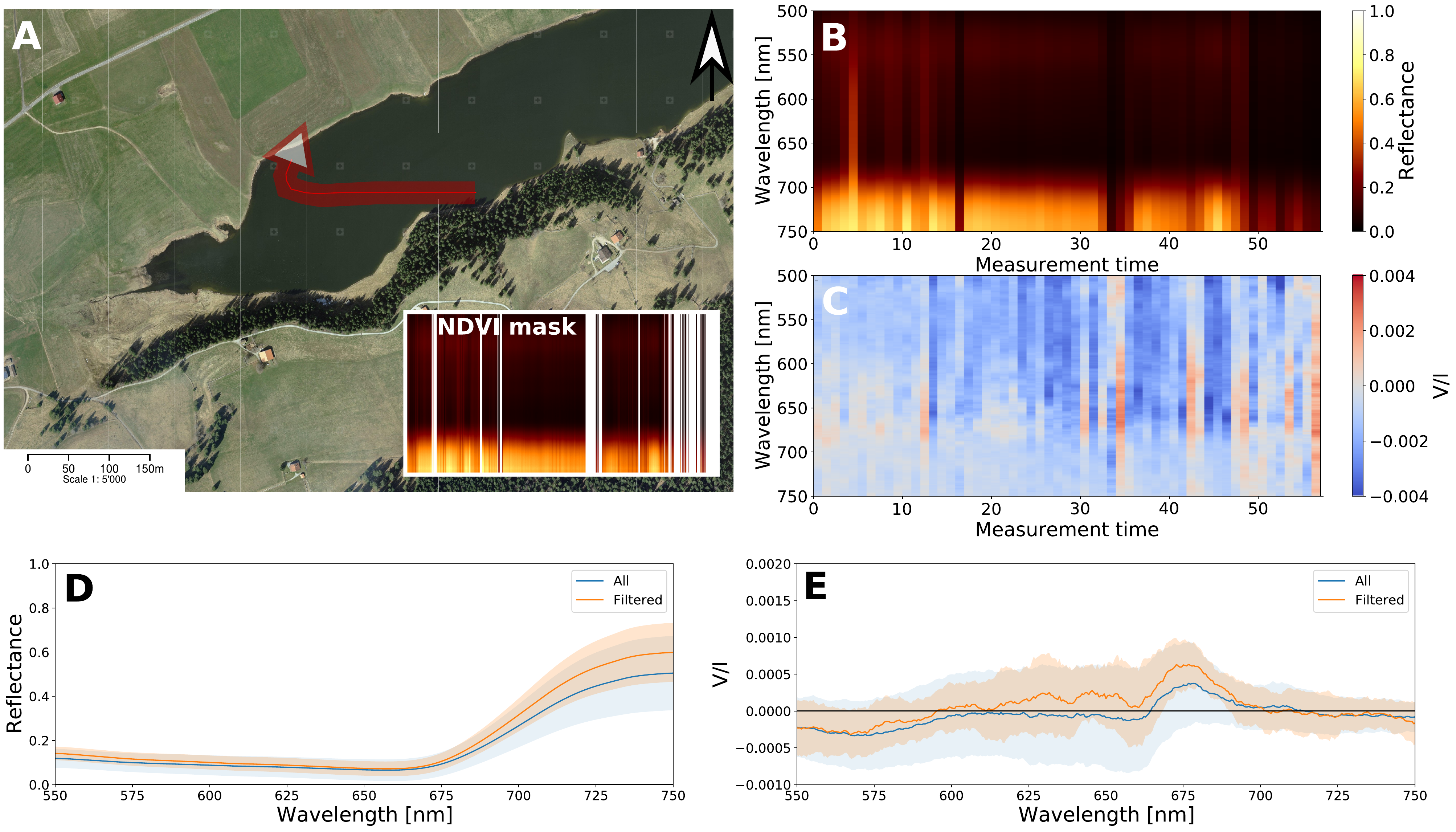}
        \centering
        \caption{\textbf{A}: One of the measurement trajectories (in red) over the lake Lac des Taill\`eres in La Br\'evine, flown in W direction. The inset indicates the filtered classification using NDVI (same color bar and scale as \textbf{B}). Satelite image by: Swiss Federal Authorities, CNES, Spot Image, Swisstopo, NPOC. \textbf{B}: Qualitative reflectance data for the trajectory per wavelength in time [s]. \textbf{C}: V/I for the trajectory per wavelength in time [s]. \textbf{D}: Time-averaged reflectance for the whole lake and filtered NDVI. \textbf{E}: Time-averaged circular polarization for the whole lake and filtered NDVI. The shaded areas denote the standard deviation between measurements.}
        \label{fig:Lac}
\end{figure*}
A trajectory was flown over the lake Lac des Taill\`eres, which is shown in Figure \ref{fig:Lac}. A red edge is only visible along part of the trajectory (Figure \ref{fig:Lac} \textbf{B} ) and, for instance, is absent after 49 seconds. The lack of a visible red edge does not necessarily indicate the absence of photosynthetic organisms but could also be  the effect of the lower total photon count, which might be related to the lake depth. The average fractional circular polarization for the whole scene as well as for the filtered scene (for low NDVI and/or unusual high and low counts) are given in Figure \ref{fig:Lac} \textbf{E}. 
 
\subsection{Validation}

\begin{figure*}[!htb]
        \centering 
        \includegraphics[width=0.95\textwidth]{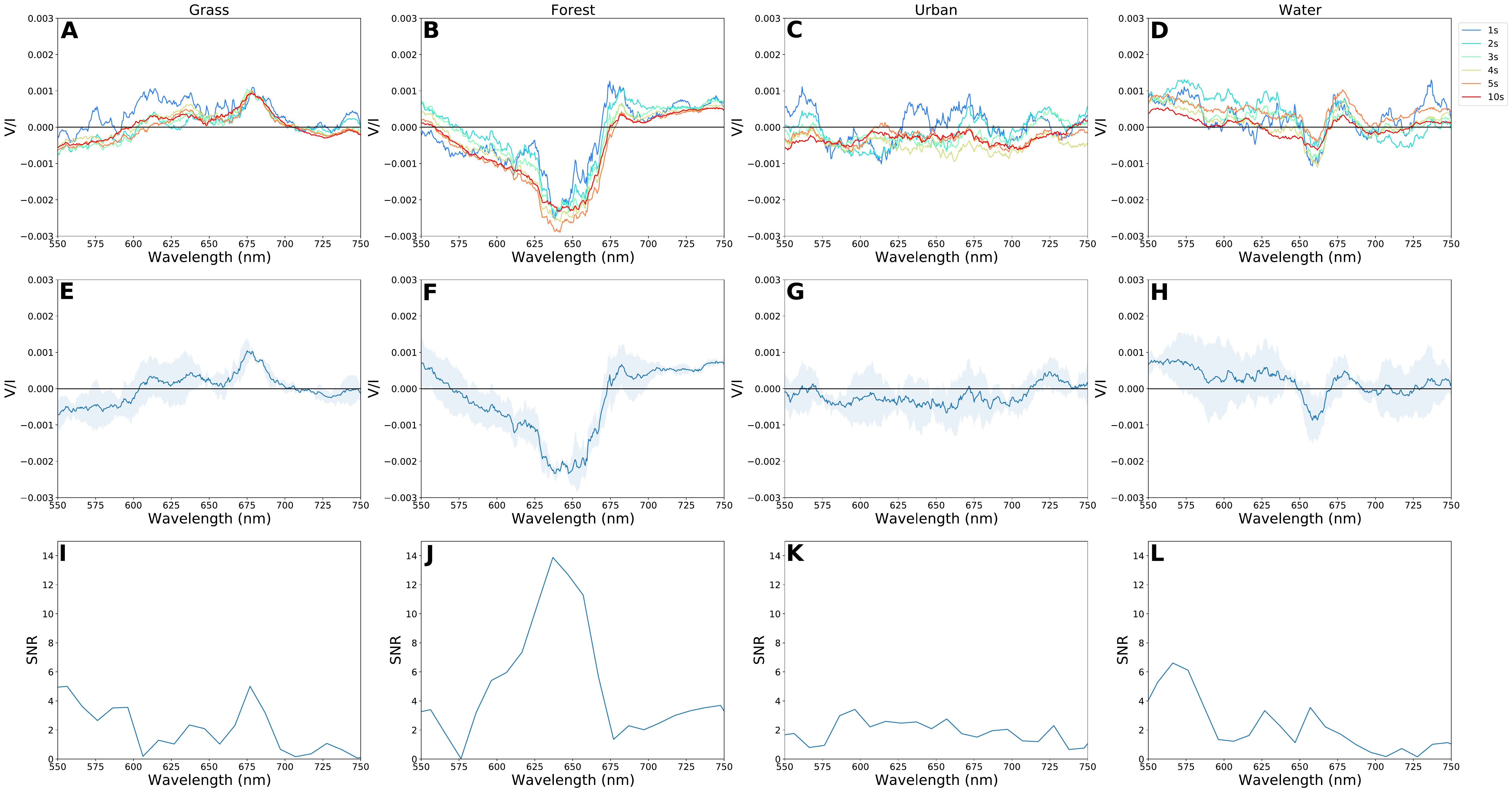}
        \caption{Circular polarimetric spectra of grass (\textbf{A}, \textbf{E}), forest (\textbf{B} and \textbf{F}) urban (\textbf{C} and \textbf{G}), and water (\textbf{D} and \textbf{H}). The upper panels show the fractional circular polarization per wavelength for different measuring time intervals. The middle panel shows the average of three subsequent one second measurements; the standard deviation between them is denoted by the shaded area and the corresponding S/N is shown in \textbf{I}, \textbf{J}, \textbf{K,} and \textbf{L}.}
        \label{fig:Time}
        
\end{figure*}

In Figure \ref{fig:Time} we show the circular polarization over shorter measurement durations and the signal-to-noise ratio (S/N) for the different landscape elements: \textbf{A}, \textbf{E,} and \textbf{I} indicate grass; \textbf{B}, \textbf{F,} and \textbf{J} represent trees; \textbf{C,} \textbf{G,} and \textbf{K} indicate urban; and \textbf{D}, \textbf{H,} and L indicate water. The upper panels show the circular polarization spectra for different time intervals: per 1, 2, 3, 4, 5, and 10 s. The center panels show the average of three subsequent one second measurements and the standard deviation between those scenes is denoted by the shaded area. The top panels show the detection S/N for the center panels using the offset substracted measurements on spectralon as sigma. In general, clear features are already present for measurements of  one second\ duration and the repeatability is high as is demonstrated by the center panels of Figure \ref{fig:Time} with a S/N  larger than 5 for vegetation.

The grass does not exhibit significant differences for measurements longer than two seconds (Figure \ref{fig:Time} \textbf{A}) with a positive band maximum $V/I_{max}=1.0\cdot10^{-3}$ at 680 nm. For tree canopies (Figure \ref{fig:Time} \textbf{B}), the variations are larger; where the measurement of shorter duration start with a positive band maximum $V/I_{max}=1.3\cdot10^{-3}$ this reduces to $V/I_{max}=3.2\cdot10^{-4}$ for 10 seconds. The measurements over the urban area (Figure \ref{fig:Time} \textbf{C}) show no clear features for the longer measurement durations. The observations of the water surface (Figure \ref{fig:Time} \textbf{D}) generally display a signal amplitude that is smaller on average than that of vegetation. A nonzero signal can, however, be observed around the chlorophyll absorbance band.  

Figure \ref{fig:Intsp} shows polarimetric measurements carried out in the laboratory on three different vegetation samples and concrete. In reflection, leaves often display two main spectral types of circular polarizance. One is characterized by the spectrum in Figure \ref{fig:Intsp} \textbf{B,} which is similar to \textbf{A} with both a positive and a negative band. Additionally, in many instances a signal with only a negative band is observed (Figure \ref{fig:Intsp} \textbf{C}). The measurements on concrete show a clear zero signal, where $V/I_{max}=2.5\cdot10^{-5}$ and $V/I_{min}=2\cdot10^{-5}$.

\section{Discussion and conclusions}
\begin{figure*}[!ht]
        \centering 
        \includegraphics[width=0.95\textwidth]{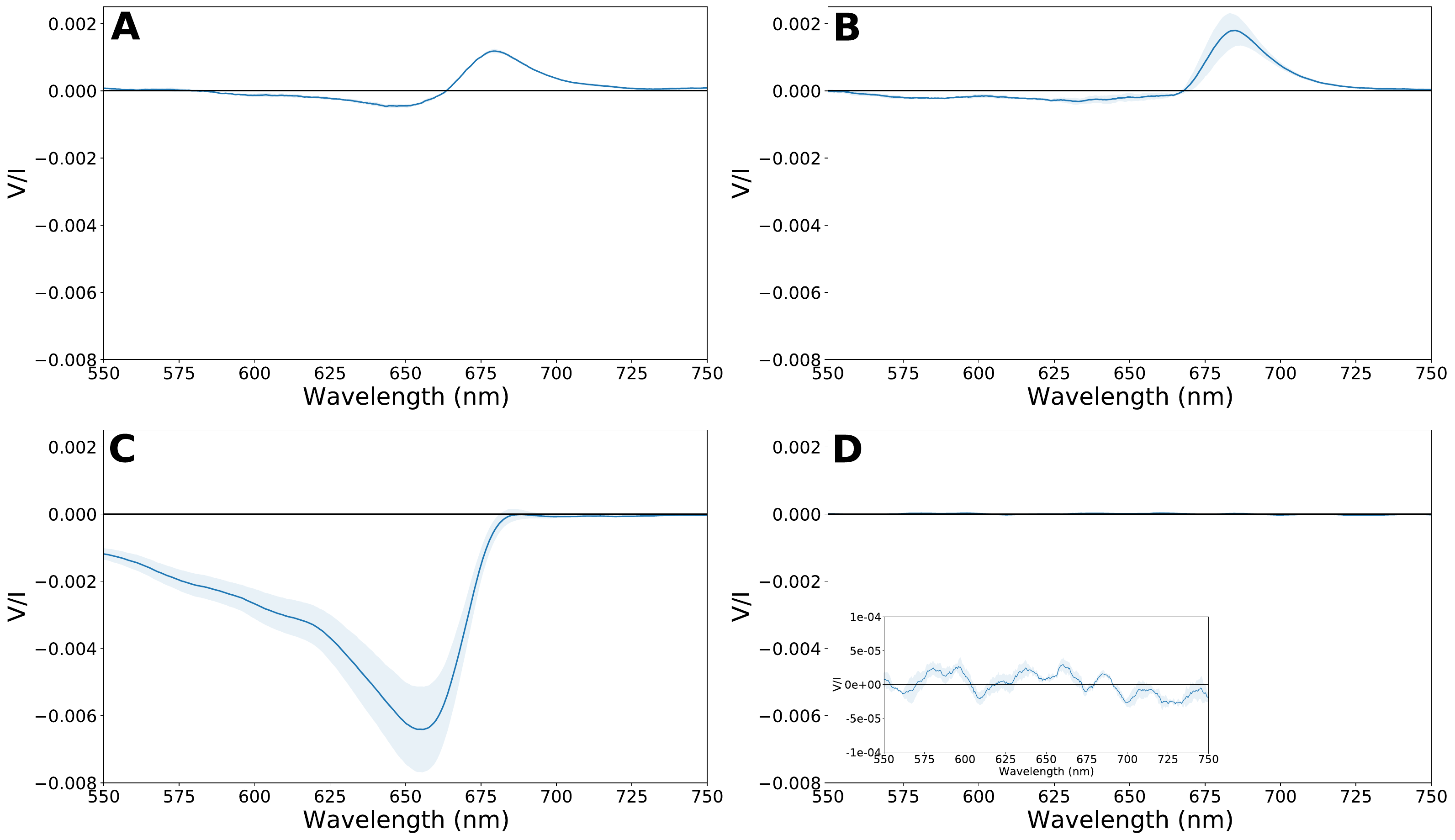}
        \caption{Circular polarimetric spectra measured (in reflection) in the laboratory of grass (\textit{Gramineae sp.}) (\textbf{A}), tree leaves,
  \textit{ Acer platanoides} (\textbf{B}),  \textit{Platanus sp.} (\textbf{C}), and concrete (\textbf{D}). The inset provides a close-up of the same measurement. The shaded area denotes the standard deviation, n=3.}
        \label{fig:Intsp}
\end{figure*}

We show for the first time, to our knowledge, the possibility of acquiring high-quality circular polarization spectra from a fast-moving aerial platform with a total measurement duration down to one second. We were able to distinguish vegetation from abiotic surfaces and demonstrate for the first time the possibility of detecting life in lakes through spectropolarimetry, underlining the potential use of life's circular polarization as an agnostic biosignature for the detection of life beyond Earth.

With the average of three subsequent one-second duration measurements (Figure \ref{fig:Time}), we were able to distinguish between the forest, grass, and urban areas with a S/N of almost 14 for forests and 5 for grass. While our measurements over water bodies display a spectropolarimetric signal that is indicative of the presence of phototrophic organisms, the signal is low in amplitude. The noise in the urban area measurements is relatively large compared to the measurements on concrete in the laboratory (shown in Figure \ref{fig:Intsp} \textbf{D)}. This is likely a consequence of the large number of different geometries (e.g., by roofs and walls), the presence of objects with high linear polarization (e.g.,  windows and cars) and the possible presence of vegetation (i.e., in gardens and along roads). Nonetheless, the spectrum is essentially free from any significant spectral features compared to areas covered by phototrophic life. 

The spectral characteristics of the measurements in this work are qualitatively similar in shape to those obtained in transmission in the laboratory, that is, showing negative and positive bands that are characteristic of the chloroplast thylakoid membranes of plants. These two bands are termed ``psi-type'' (polymer- and salt-induced) circular polarizance or dichroism and they are associated with the long-range chiral order of chlorophyll molecules in ordered macroarrays of the chlorophyll-protein complexes in the membrane \citep{Garab2009}. The two bands have different physical origins, where the negative band is associated preferentially with the stacking of the thylakoid membranes inside the chloroplasts, whereas the positive band is mainly associated with the lateral organization of the chiral macro-domains \citep{Lambrev2019}. We note that although it appears that the negative band is absent in for instance the grass, this is likely due to a slight offset in the zero point. The positive band is often less well characterized in measurements taken at a distance (cf. \citep{Patty2019}) and is also characteristic for certain trees such as the \textit{Platanus} shown in Figure \ref{fig:Intsp} \textbf{C}. We observed similar variations for some of the forest trajectories, shown in Figure \ref{fig:Time} \textbf{B}, where the initially larger positive band decreases in magnitude with more captured frames. In other trajectories this effect was less pronounced, which was also observed for different scenes in \cite{Patty2019}. Similar spectral phenomena occur around the veins of leaves, where it was suggested to have been caused by the orientation of the chloroplasts \citep{Patty2018b}. It is at this point unclear what the underlying mechanisms are for the results on whole leaf canopies and how this might vary between different species. 

We also would like to emphasize that it might be possible that, to some extent, the phase angles involved play a role in this phenomenon; {\color{red}  we did not accurately record this during the flight campaign.} Additionally, as of yet, no systematic experimental study exists on the effect of incidence and reflection angles on vegetation polarizance. Such a study would additionally aid in the development of an (exo)planetary model with realistic circular polarization surface components \citep{Groot2020}. 

One of the key limiting factors in terms of measuring time and sensitivity is the accurate scrambling of linear polarization to prevent linear-to-circular polarization crosstalk ($Q,U \rightarrow V$ components). Especially from a fast-moving platform, large fluctuations on a short timescale can be expected. The data (see also Figure \ref{fig:Time}) indicate that a one-second measurement time generally ensures that the scrambling of the linear polarization allows for sub-sequential subtraction. It would be interesting to observe the variability in linear polarizance originating from these different landscape elements. In terms of remote sensing it has been suggested that linear polarization offers no additional information compared to the scalar reflectance \citep{Peltoniemi2015}, although it has been demonstrated that linear polarization  can inform plant physiology as well \citep{Vanderbilt2017,Vanderbilt2019}. We intend to include linear polarization observations in future flight campaigns. Capturing linear and circular polarization simultaneously, however, would require an additional second polarimeter or a full-Stokes polarimeter that is capable of acquiring the polarization on a single data frame (see, e.g.,  \citep{Snik2019, Sparks2019, Keller2020}).

While subsequent measurements are often very repeatable (as shown in Figure \ref{fig:Time}), it is shown in Figure \ref{fig:Sat} \textbf{C} that in some cases no clear signal is measured while generally a clear red edge is visible for the whole scene with the exception of the roads (Figure \ref{fig:Sat} \textbf{B}). It is possible that the lack of clear $V/I$ signals in this case might be due to local variations in the grass physiology. The measurements were taken at the end of summer and it is  visible in \ref{fig:Helipic} \textbf{A} that there are various dry patches in the grass. It has been demonstrated before that drought can greatly decrease the circular polarizance of the grass, while it does not decrease the red edge in a similar manner \citep{Patty2017}. Nonetheless, for this particular campaign we did not take ground $V/I$ observations that could elucidate.

Sensitive spectropolarimetry could prove to be a very valuable tool for the remote sensing of vegetation and other photosynthetic organisms on Earth. The psi-type circular polarizance of vegetation in particular is a sensitive indicator of the photosynthetic membrane macro-organization, which can show changes and variations as a result of various stress factors \citep{Lambrev2019}. It has additionally been demonstrated that various species display different spectral characteristics. As such, circular polarimetry could be used, for instance, to measure deforestation and diseases in forests and it might, in addition, be possible to distinguish between tree species and to monitor evasive species. The first results on water bodies in this study look promising. It could be valuable to implement circular polarization in the monitoring of toxic algal blooms, of coral reefs, and the effects of acidification thereon. Additionally, it could aid as a sensitive indicator of pollution, such as by spills, by measuring the polarization response of phytoplankton.

{\color{red}While promising in terms of biosignature exclusivity, the magnitude of the circular polarization signal is low and typically much less than 1\% in the field. As such the detection of similar circular polarization signals originating from exoplanets will be challenging. Additionally, in a planetary disk average it can be expected that the signal is further diluted by surfaces with a reflectance that is higher than those creating the circular polarization. This is also the case for clouds, which are highly reflective and can hide the circular polarization signal and other surface biosignatures such as the red edge. In general, however, the atmospheric circular polarization signal is very low and allows for the detection of circular polarization signals originating from surface biological sources \citep{Rossi2018}. It is unlikely that the next generation of telescopes could measure circular polarization signals with a similar magnitude originating from Earth-like planets around solar-type stars. In theory, such signals could however be detected with only 100 hours of observations on a terrestrial super Earth in close orbit (0.05 AU) to the star based on the instrumental performance of ground-based telescopes and the precision of POLLUX \citep{Sparks2021}. In the end, a confident detection of extraterrestrial life will naturally have to consist of a consistent range of spectroscopic biosignatures \citep{Seager2014}, of which the detection of circular polarization signals will be very strong evidence of homochirality and hence life.}

We have successfully demonstrated the possibility of using sensitive circular spectropolarimetry from a fast-moving aerial platform. An important next step will be to measure the circular polarization of the Earth surface from an orbiter such as from the International Space Station. Even with a one second measurement time; this would result in roughly a 6-7 km resolution and will provide valuable information for future polarization remote sensing and will also provide crucial input on the potential use of life's circular polarization as an agnostic biosignature for the detection of life beyond Earth.

\section*{Acknowledgments}
This work has been carried out within the framework of the National Centre of Competence in Research (NCCR), PlanetS, supported by the Swiss National Science Foundation (SNSF). We thank Jean-Claude Pointet for flying the helicopter. The research of FS leading to these results has received funding from the European Research Council under ERC Starting Grant agreement 678194 FALCONER.

\bibliographystyle{aa}
\bibliography{Alles}

\end{document}